\documentclass[a4paper,11pt]{article}

\usepackage{jheppub}
\usepackage[T1]{fontenc}
\usepackage{graphicx}
\usepackage{hyperref}
\usepackage{amsmath}
\usepackage{epsfig}
\usepackage{subfig}
\usepackage{xcolor}
\usepackage{natbib}
\usepackage{bm}

\def\be{\begin{equation}}
\def\ee{\end{equation}}
\def\bea{\begin{eqnarray}}
\def\eea{\end{eqnarray}}
\def\NO{\nonumber}

\title{Hadronic decay of exotic mesons consisting of four charm quarks}

\author[a,b]{Hong-Fei Zhang,}
\author[b]{Xue-Mei Mo,}
\author[c]{Yu-Peng Yan}
\affiliation[a]{College of Big Data Statistics, Guizhou University of Finance and Economics, Guiyang, 550025, China}
\affiliation[b]{Research Centre of Big-data Corpus \& Language Projects, School of Foreign Languages, Guizhou University of Finance and Economics, Guiyang, 550025, China}
\affiliation[c]{School of Physics and Center of Excellence in High Energy Physics $\mathrm{\&}$ Astrophysics, Suranaree University of Technology, Nakhon Ratchasima 30000, Thailand}
\emailAdd{shckm2686@163.com}
\emailAdd{dmtmmu@163.com}
\emailAdd{yupeng@sut.ac.th}

\abstract{
In 2020, a di-$J/\psi$ resonance centered at around 6.9 $\mathrm{GeV}$ was discovered by the LHCb Collaboration, and was later confirmed by the ATLAS and CMS Collaborations.
In addition to the 6.9-$\mathrm{GeV}$ resonance, its low energy counterpart centered at around 6552 $\mathrm{MeV}$ was also observed by the CMS Collaboration.
In this paper, we calculate the inclusive decay width of such exotic mesons into light hadrons and into open charms, respectively,
assuming them as either genuine tetraquarks or molecules composed of two $J/\psi$'s.
Since the spins of these resonances have not yet been experimentally measured, regarding them as S-wave bound states,
we consider all possible spin configurations, namely spin-0 and spin-2, that decay exclusively into double $J/\psi$,
and find that the ratio ($\mathcal{R}$) of the decay width for the light-hadron final states to that for the charmed final states differs significantly among different spin configurations,
which can help to explore the nature of the observed di-$J/\psi$ resonances and fix the mixing angle for the $J^{PC}=0^{++}$ tetraquark.
}

\begin{document}

\maketitle
\bibliographystyle{JHEP}

\section{Introduction\label{sec:introduction}}

In 2020, the LHCb Collaboration observed a di-$J/\psi$ resonance at around 6.9 $\mathrm{GeV}$,
which in most of the literature was considered a state consisting of two pairs of charm-anticharm quarks.
This resonance was later confirmed by the ATLAS and CMS Collaborations,
the latter of which in addition discovered a new resonance centered at around 6552 $\mathrm{MeV}$.
These new states opened a new window for exploring the physics underlying exotic hadrons.
First of All, the involved particles are heavy enough so that the perturbation theory works.
Further, for such hadrons composed solely of charm quarks\footnote{
We regard them as tetraquarks in this paper, and use the term genuine tetraquarks or compact tetraquarks referring to those in which the constituent quarks are compactly bound.},
one can impose a natural principle to discriminate in theory between compact tetraquarks and hadronic molecules:
the wave functions of compact tetraquarks should be asymmetric with respect to exchanging the two quarks (antiquarks),
while those of hadronic molecules should be symmetric with respect to exchanging the two hadrons,
which may lead to much difference between the two in production and decay processes.

To date, the mass spectra of fully-heavy tetraquarks have been extensively studied using various models and techniques in the literature
\cite{Chao:1980dv, Ader:1981db, Zouzou:1986qh, Heller:1985cb, Lloyd:2003yc, Barnea:2006sd, Berezhnoy:2011xn, Wu:2016vtq, Chen:2016jxd, Karliner:2016zzc, Bai:2016int, Wang:2017jtz, Anwar:2017toa,
Debastiani:2017msn, Esposito:2018cwh, Wang:2018poa, Liu:2019zuc, Wang:2019rdo, Chen:2020lgj, Lundhammar:2020xvw, Bedolla:2019zwg, Deng:2020iqw, Yang:2020rih, Wang:2020ols, Jin:2020jfc, Lu:2020cns,
Albuquerque:2020hio, Sonnenschein:2020nwn, Giron:2020wpx, Wang:2020wrp, Yang:2020atz, Karliner:2020dta, Wang:2020dlo, Zhao:2020nwy, Gordillo:2020sgc, Faustov:2020qfm, Weng:2020jao, Zhang:2020xtb, Zhu:2020xni,
Zhao:2020jvl, Albuquerque:2021erv, Faustov:2021hjs, Ke:2021iyh, Yang:2021hrb, Mutuk:2021hmi, Li:2021ygk, Majarshin:2021hex, Pal:2021gkr, Wang:2021mma, Asadi:2021ids, Kuang:2022vdy, Wu:2022qwd,
Liang:2022rew, An:2022qpt, Zhang:2022qtp, Faustov:2022mvs, abu-shady:2022nwi, Agaev:2023gaq, Agaev:2023ruu, Kuchta:2023anj},
most of which assume the di-$J/\psi$ resonances as fully-charm compact tetraquark states.
However, the identities of the observed resonances are still under debate;
many works have been dedicated to the interpretation of them
\cite{Richard:2020hdw, Chao:2020dml, Dong:2020nwy, Cao:2020gul, Gong:2020bmg, Wan:2020fsk, Liang:2021fzr, Dong:2021lkh, Zhuang:2021pci, Wang:2022jmb, Zhou:2022xpd, Lu:2023ccs, Kuang:2023vac},
most considering these resonances as effects of two charmonia produced simultaneously.
We call such effects, regardless of whether the two charmonia are bound, a molecule-like state.

Unfortunately, there still does not exist an approach that can achieve high precision prediction of hadron masses,
as a result, one cannot expect to make definite conclusion of the identity of the di-$J/\psi$ resonances based only on the calculation of their masses.
Therefore, the investigation of the production and decay of the candidates becomes crucial.
There have been only a few references studying the production
\cite{Berezhnoy:2011xn, Wang:2020gmd, Maciula:2020wri, Zhang:2020hoh, Feng:2020riv, Feng:2020qee, Zhu:2020xni, Zhao:2020nwy, Goncalves:2021ytq, Szczurek:2022orx, Feng:2023agq}
and decay~\cite{Chao:1980dv, Chen:2020xwe, Yang:2021hrb, Chen:2022sbf, Biloshytskyi:2022dmo, Agaev:2023gaq, Agaev:2023ruu, Sang:2023ncm} of fully-heavy tetraquarks.
Worthy of mentioning, References~\cite{Zhang:2020hoh, Feng:2020riv} for the first time extended nonrelativistic QCD~\cite{Bodwin:1994jh} to the evaluation of fully-heavy tetraquark production.
Reference~\cite{Zhang:2020hoh} carried out the first calculation of the transverse momentum distribution of the fully-heavy tetraquark (including molecule composed of two $J/\psi$'s) hadroproduction,
in accord with the experimental condition for discovering the di-$J/\psi$ resonances.
Their results were partially confirmed a few year later by Reference~\cite{Feng:2023agq}.
These results showed that the production cross section for the spin-2 fully-heavy tetraquark are at least one order of magnitude greater than that for its spin-0 counterparts,
which favours spin-2 nature of the observed di-$J/\psi$ resonances.
Note that this information is not easily accessible through the mass spectra.

Akin to the production cases, theoretical study of the fully-heavy tetraquark decay widths,
especially the relative fractions for some species of final states, may also provide useful information.
By exploiting the Fierz identities, References~\cite{Chen:2020xwe, Chen:2022sbf} investigated relative branching fractions for some of the double-quarkonia final states.
In Reference~\cite{Sang:2023ncm}, the authors evaluated the decay widths of compact fully-heavy tetraquarks to light hadrons and to two photons within the frameworks of nonrelativistic QCD.
In this paper, we calculate the inclusive decay width of fully-charm tetraquarks (including molecule-like states) into both light hadrons and open charms,
as well as the ratio of one to the other.
We will demonstrate that this ratio can provide crucial information for the nature of the observed resonances,
as long as it could be measured to a specific precision.
These results also applies to the decay of fully-bottom tetraquarks once having the masses and wave functions involved replaced by the bottom counterparts.

This paper is organised as follows.
The calculation framework will be elaborated in Section~\ref{sec:framework}.
In Section~\ref{sec:result}, the results will be presented and their phenomenological indications will be discussed.
Section~\ref{sec:summary} is a concluding remark.

\section{Theoretical Framework\label{sec:framework}}

\subsection{Nonrelativistic QCD factorization for fully-charm tetraquark decay}

Within the nonrelativistic QCD framework~\cite{Bodwin:1994jh},
the decay width of a fully-charm multiquark state in principle can be factorized as
\bea
\Gamma(H\rightarrow X)=\sum_n\langle\mathcal{O}^H(n)\rangle f(n\rightarrow X),
\eea
where $X$ denotes the final states,
$n$ stands for the intermediate state of the constituent charm quarks with specific quantum numbers,
$\langle\mathcal{O}^H(n)\rangle$ is the long-distance matrix element describing the possibility of finding the intermediate state $n$ inside the hadron $H$,
and $f(n\rightarrow X)$ is the short-distance coefficient, a quantity that can be evaluated perturbatively,
describing the decay of the intermediate state $n$ into the final states $X$.
$n$ runs over all possible quantum numbers which contribute to the decay width,
given the restrictions required by the conservation rules.
Each long-distance matrix element is assigned an order in $v$,
the typical relative velocity of the constituent quarks inside the hadron $H$.
For charmonia, $v^2\approx0.3$, while for bottomonia, $v^2\approx0.1$.
However, for a multiquark state consisting of more than two constituent quarks,
more than one velocity scales might arise, which adds to the complexity of the long-distance matrix elements.

In this paper, we are concerned only with colour-singlet contributions,
i.e. when the quantum numbers of $n$ and $H$ coincide, for the following reasons.
First of all, the contributions from colour-multiplet states are suppressed, especially for decay processes.
Moreover, the long-distance matrix elements for colour-multiplet states have not yet been determined due to lack of data.

Within the colour-singlet model, the decay width of an S-wave fully-charm tetraquark ($T$) can be written as~\cite{Zhang:2020hoh}
\bea
\Gamma(T\rightarrow X)=|\Psi(0)|^2\hat{\Gamma}(T\rightarrow X),
\eea
where $\Psi$ is the spacial wave function of $T$,
and $\hat{\Gamma}$ is the appropriately normalized short-distance coefficient.
Note that in the short-distance coefficient, we denote the $cc\bar{c}\bar{c}$ intermediate state by the same letter ($T$) as that to denote the tetraquark.
$|\Psi(0)|^2$ stands for the probability density of finding all the four constituent quarks at the same position.

Although nonrelativistic QCD provides complete framework to separate the long-distance and short-distance effects,
it does not tell how the colour-singlet matrix elements are related to the corresponding wave functions,
which in any case depends on the simplified model of how the tetraquark state is built in terms of those of its constituents.

In the rest of this section, we present the formalism of calculating the decay widths of molecule-like states and compact tetraquarks in terms of tetraquark wavefunctions,
which are constructed conventionally as superpositions of products of the constituent quarks.

\subsubsection{Molecule-like states}

The extension of nonrelativistic QCD to calculating the decay width of a molecule-like state composed of two $J/\psi$'s is quite straightford.

The the ladder operator to create a molecule-like state can be formally written as
\bea
&&a_{T_M}^\dagger(P,J,J_z)=\sum_{\lambda_1\lambda_2}C_{JJ_z}^{\lambda_1\lambda_2}\int\frac{\mathrm{d}^3p_1}{(2\pi)^32p_1^0}\frac{\mathrm{d}^3p_2}{(2\pi)^32p_2^0} \NO \\
&&~~~~\times(2\pi)^32P^0\delta^3(P-p_1-p_2)\tilde{\psi}_M(P,p_1,p_2)a_{J/\psi}^\dagger(p_1,\lambda_1)a_{J/\psi}^\dagger(p_2,\lambda_2), \label{eqn:moleculestate}
\eea
where the $J/\psi$ creation operators are normalized such that
\bea
\Big[a_{J/\psi}(p',\lambda'),a_{J/\psi}^\dagger(p,\lambda)\Big]=(2\pi)^32p^0\delta^3(\bm{p}-\bm{p'})\delta_{\lambda\lambda'}.
\eea
The spacial wave function $\tilde{\psi}_M(P,p_1,p_2)$ depends solely on the difference between $p_1$ and $p_2$,
thus we rewrite it as $\tilde{\psi}_M(q)$, where $q=(p_2-p_1)/2$.
Denoting the $J/\psi$ and tetraquark mass as $m$ and $M$, respectively, we have
\bea
\frac{\mathrm{d}^3p_1}{(2\pi)^32p_1^0}\frac{\mathrm{d}^3p_2}{(2\pi)^32p_2^0}&=&\frac{\mathrm{d}^4p_1}{(2\pi)^3}\frac{\mathrm{d}^4p_2}{(2\pi)^3}\delta(p_1^2-m^2)\delta(p_2^2-m^2)\theta(p_1^0)\theta(p_2^0) \NO \\
&=&\frac{\mathrm{d}^4P}{(2\pi)^3}\frac{\mathrm{d}^4q}{(2\pi)^3}\delta(P^2-4(m^2-q^2))\delta(\frac{P}{2}\cdot q)\theta(\frac{P^0}{2}+q^0)\theta(\frac{P^0}{2}-q^0) \NO \\
&=&\frac{\mathrm{d}^3P}{(2\pi)^32P^0}\frac{\mathrm{d}^3q}{(2\pi)^3}\frac{2P^0}{(P^0)^2-4(q^0)^2},
\eea
employing which, Equation~(\ref{eqn:moleculestate}) reduces to
\bea
&&a_{T_M}^\dagger(P,J,J_z)=\sum_{\lambda_1\lambda_2}C_{JJ_z}^{\lambda_1\lambda_2}\int\frac{\mathrm{d}^3q}{(2\pi)^3}\frac{2P_0}{(P^0)^2-4(q^0)^2} \NO \\
&&~~~~\times\tilde{\psi}_M(q)a_{J/\psi}^\dagger(\frac{P}{2}+q,\lambda_1)a_{J/\psi}^\dagger(\frac{P}{2}-q,\lambda_2), \label{eqn:ms}
\eea
with
\bea
P^0=\sqrt{\bm{P}^2+M^2},~~~~~~~~q^0=\sqrt{m^2+\bm{q}^2-\frac{M^2}{4}}.
\eea
Note that in tetraquark rest frame, $P^0=M$, $q^0=0$, and $|\bm{q}|=\sqrt{M^2/4-m^2}$.

We adopt the same normalization for the molecule-like state vector, namely
\bea
\Big[a_{T_M}(P',J',J_z'),a_{T_M}^\dagger(P,J,J_z)\Big]=(2\pi)^32P^0\delta^3(\bm{P}-\bm{P'})\delta_{JJ'}\delta_{J_zJ_z'},
\eea
which, with Equation~(\ref{eqn:ms}) substituted, leads to
\bea
\int\frac{\mathrm{d}^3q}{(2\pi)^3}\frac{2P^0}{(P^0)^2-4(q^0)^2}2\big|\tilde{\psi}_M(q)\big|^2=1. \label{eqn:normalization}
\eea
The factor of 2 in front of the squared wave function arises from the symmetry of the tetraquark wave function with respect to exchanging the two $J/\psi$'s.
In the tetraquark rest frame, the normalization equation becomes
\bea
\int\frac{\mathrm{d}^3q}{(2\pi)^3}\frac{2}{M}2\big|\tilde{\psi}_M(\bm{q})\big|^2=1.
\eea

The Feynman amplitude for the decay of a molecule-like state into the final states $X$ can thus be written as
\bea
&&\langle X|T_M(P,J,J_z)\rangle=\langle X|a_{T_M}^\dagger(P,J,J_z)|0\rangle \NO \\
&&~~~~~~=\sum_{\lambda_1\lambda_2}C_{JJ_z}^{\lambda_1\lambda_2}\int\frac{\mathrm{d}^3q}{(2\pi)^3}\frac{2P_0}{(P^0)^2-4(q^0)^2}\tilde{\psi}_M(q) \NO \\
&&~~~~~~\times\langle X|J/\psi(\frac{P}{2}+q,\lambda_1)J/\psi(\frac{P}{2}-q,\lambda_2)\rangle.
\eea
So far, we have retained the relativistic nature of the amplitude.
In order to further simplify the Feynman amplitude, we carry out the integration in tetraquark rest frame,
and the explicit Lorentz invariance of the amplitude will be restored afterwards.
The integral can be easily evaluated by expanding the amplitude for double $J/\psi$ to $X$ as a linear combintation of orbital angular momentum eigenstates,
all terms of which vanish except for that has the same orbital angular momentum as the spacial wave function $\tilde{\psi}_M$.

For S-wave tetraquark production, the amplitude in tetraquark rest frame reduces to
\bea
&&\langle X|T_M(P,J,J_z)\rangle=\sum_{\lambda_1\lambda_2}C_{JJ_z}^{\lambda_1\lambda_2}
\int\frac{\mathrm{d}^3q}{(2\pi)^3}\tilde{\psi}_M(\bm{q})\langle X|J/\psi(\frac{P}{2},\lambda_1)J/\psi(\frac{P}{2},\lambda_2)\rangle \NO \\
&&~~~~=\sum_{\lambda_1\lambda_2}C_{JJ_z}^{\lambda_1\lambda_2}\sqrt{\frac{2}{M}}\frac{1}{\sqrt{2}}
\psi_M(\bm{0})\langle X|J/\psi(\frac{P}{2},\lambda_1)J/\psi(\frac{P}{2},\lambda_2)\rangle, \label{eqn:msamp}
\eea
where $\psi_M$ is the normalized spacial wave function with coordinate representation.
Equation~(\ref{eqn:msamp}) is manifestly Lorentz invariant, which implies its validity in arbitrary frame of reference
as long as the value of the wave function at the origin is evaluated in the tetraquark rest frame.

The $J/\psi$ decay amplitude is proportional to its polarization vector, namely
\bea
\langle X|J/\psi(\frac{P}{2},\lambda_1)J/\psi(\frac{P}{2},\lambda_2)\rangle=
\langle X|J/\psi(\frac{P}{2})J/\psi(\frac{P}{2})\rangle_{\mu\nu}\epsilon_{\lambda_1}^\mu(\frac{P}{2})\epsilon_{\lambda_2}^\nu(\frac{P}{2}),
\eea
which can further simplify the amplitude,
\bea
\langle X|T_M(P,J,J_z)\rangle=\sqrt{\frac{2}{M}}\frac{1}{\sqrt{2}}\psi_M(\bm{0})\langle X|J/\psi(\frac{P}{2})J/\psi(\frac{P}{2})\rangle_{\mu\nu}\epsilon_{JJ_z}^{\mu\nu},
\eea
where
\bea
\epsilon_{JJ_z}^{\mu\nu}=\sum_{\lambda_1\lambda_2}C_{JJ_z}^{\lambda_1\lambda_2}\epsilon_{\lambda_1}^\mu(\frac{P}{2})\epsilon_{\lambda_2}^\nu(\frac{P}{2}).
\eea
For $J=0$,
\bea
\epsilon_{00}^{\mu\nu}=\sqrt{\frac{1}{3}}\Pi^{\mu\nu},
\eea
and for $J=2$,
\bea
\sum_{J_z}\epsilon_{2J_z}^{\mu\nu}(\epsilon_{2J_z}^{\mu'\nu'})^*=\frac{1}{2}(\Pi^{\mu\mu'}\Pi^{\nu\nu'}+\Pi^{\mu\nu'}\Pi^{\nu\mu'})-\frac{1}{3}\Pi^{\mu\nu}\Pi^{\mu'\nu'},
\eea
where
\bea
\Pi^{\mu\nu}=-g^{\mu\nu}+\frac{P^\mu P^\nu}{M^2}.
\eea

\subsubsection{Compact tetraquarks}

The primary difference between a compact tetraquark and a molecule-like tetraquark in nonrelativistic QCD framework
lies in the fact that a compact tetraquark is composed of a pair of identical quarks and a pair of identical antiquarks,
while a molecule-like tetraquark is composed of two quarkonia.
Under this hypothesis, the wave function of a compact tetraquark should be asymmetric while exchanging the two constituent quarks (antiquarks).
Another difference between the two sorts of tetraquarks is that the wave function of a compact tetraquark has colour degree of freedom
while it is not the case for a molecule-like state.

For the cases studied in this paper, say S-wave tetraquarks that exclusively decay into two $J/\psi$'s,
the colour wave function of a compact tetraquark can be
\bea
G^+_{ijkl}=\frac{1}{\sqrt{24}}(\delta_{ik}\delta_{jl}+\delta_{il}\delta_{jk}),
\eea
or
\bea
G^-_{ijkl}=\frac{1}{\sqrt{12}}(\delta_{ik}\delta_{jl}-\delta_{il}\delta_{jk}),
\eea
of which the former is referred to as symmetric or colour-sextuplet and the latter is referred to as asymmetric or colour-triplet.
Here, $i$, $j$, $k$, and $l$ are the colour indices for the two $c$ quarks and the two $\bar{c}$ quarks, respectively.

In order to make the wave function asymmetric with respect to exchanging the two quarks (antiquarks),
the symmetric colour wave function should be associated with spin-singlet wave functions of the $cc$ and $\bar{c}\bar{c}$ systems,
while its asymmetric counterpart with spin-triplet ones.

The spacial wave function of a compact tetraquark with momentum representation depends on three independent momenta,
the relative momentum between the two quarks, $q_1=(p_1-p_2)/2$, that between the two antiquarks, $q_2=(p_3-p_4)/2$,
and that between the quark system and the antiquark system, $q=(p1+p2-p3-p4)/2$, namely $\tilde{\Psi}_C(q,q_1,q_2)$,
where $p_1$, $p_2$, $p_3$, and $p_4$ are momenta of the two quarks and two antiquarks, respectively.

Similar to the procedure of deriving the amplitude for molecule-like tetraquark decay,
the decay amplitude for a compact tetraquark can be simplified as
\bea
&&\langle X|T_C(P,J,J_z,a)\rangle=\sqrt{\frac{2}{M}}\left(\sqrt{\frac{2}{m_{cc}}}\frac{1}{\sqrt{2}}\right)
\left(\sqrt{\frac{2}{m_{\bar{c}\bar{c}}}}\frac{1}{\sqrt{2}}\right)\sum_{\lambda_1\lambda_2}\Psi_C(\bm{0})C_{JJ_z}^{\lambda_1\lambda_2} \NO \\
&&\times\sum_{\sigma_1\sigma_2\sigma_3\sigma_4}C_{\lambda\lambda_1}^{\sigma_1\sigma_2}C_{\lambda\lambda_2}^{\sigma_3\sigma_4}
\sum_{ijkl}G^a_{ijkl}\langle X|c(\frac{P}{4},\sigma_1,i)c(\frac{P}{4},\sigma_2,j)\bar{c}(\frac{P}{4},\sigma_3,k)\bar{c}(\frac{P}{4},\sigma_4,l)\rangle,
\eea
where $\Psi_C$ is the normalized spacial wave function of the compact tetraquark with coordinate representation,
$\sigma_1$, $\sigma_2$, $\sigma_3$, and $\sigma_4$ are spins of the two quarks and two antiquarks, respectively,
$m_{cc}$ ($m_{\bar{c}\bar{c}}$) is invariant mass of the $cc$ ($\bar{c}\bar{c}$) system,
and $a$ is a sign to designate the associated colour function.
For $a=+$, $G^+$ is associated and $\lambda=0$, while for $a=-$, $G^-$ is associated and $\lambda=1$.
It is reasonable to make an approximation that $m_{cc}\approx m_{\bar{c}\bar{c}}\approx M/2$.

In order to apply the spin projection, we need to reverse the Fermion lines attaching the charm quark with momentum $p_2$ exploiting the following equation,
\bea
\overline{v}_i^\alpha(p')(k\!\!\!\slash_1+m_1)\ldots(k\!\!\!\slash_n+m_n)u_j^\beta(p_2)
=\overline{v}_j^\beta(p_2)(-k\!\!\!\slash_n+m_n)\ldots(-k\!\!\!\slash_1+m_1)u_i^\alpha(p'). \label{eqn:CPT}
\eea
If the C-Parity of either of the two constituent antiquarks has not been reversed even after Equation~(\ref{eqn:CPT}) has been implemented\footnote{
This will happen when evaluating the decay of compact tetraquarks to $c\bar{c}$ final states.},
one need to apply similar operation for the Fermion line that attaches a constituent anticharm quark while does not attach the charm quark with momentum $p_1$, namely
\bea
&&\overline{v}_i^\alpha(p_{3/4})(k\!\!\!\slash_1+m_1)\ldots(k\!\!\!\slash_n+m_n)u_j^\beta(p') \NO \\
&&~~~~~~=\overline{v}_j^\beta(p')(-k\!\!\!\slash_n+m_n)\ldots(-k\!\!\!\slash_1+m_1)u_i^\alpha(p_{3/4}).
\eea

Having carried out the above mentioned operations,
the spin projection equations can be implemented, namely
\bea
\sum_{\sigma_1\sigma_2}C_{\lambda\lambda_1}^{\sigma_1\sigma_2}u(p_1,\sigma_1)\overline{v}(p_2,\sigma_2)
=\frac{1}{\sqrt{2}m_{cc}}(p\!\!\!\slash_1+m_c)\gamma_{\lambda\lambda_1}(-p\!\!\!\slash_2+m_c), \label{eqn:spinproj}
\eea
where $m_c$ is the charm quark mass, and
\bea
\gamma_{00}=\gamma^5,~~~~\gamma_{1\lambda_1}=\gamma_\mu\epsilon_{\lambda_1}^\mu(p_1+p_2).
\eea
For S-wave tetraquark decay, Equation~(\ref{eqn:spinproj}) can be written in terms of physical quantities of tetraquarks, namely
\bea
\sum_{\sigma_1\sigma_2}C_{\lambda\lambda_1}^{\sigma_1\sigma_2}u(p_1,\sigma_1)\overline{v}(p_2,\sigma_2)
=\frac{1}{\sqrt{32}}\gamma_{\lambda\lambda_1}(-P\!\!\!\!\slash+M).
\eea

Similar equations can be obtained for anticharm spinors:
\bea
&&\sum_{\sigma_3\sigma_4}C_{\lambda\lambda_2}^{\sigma_3\sigma_4}u(p_3,\sigma_3)\overline{v}(p_4,\sigma_4)=\frac{1}{\sqrt{32}}\gamma_{\lambda\lambda_2}(-P\!\!\!\!\slash+M), \NO \\
&&\sum_{\sigma_3\sigma_4}C_{\lambda\lambda_2}^{\sigma_3\sigma_4}u(p_4,\sigma_4)\overline{v}(p_3,\sigma_3)=(-1)^{\lambda+1}\frac{1}{\sqrt{32}}\gamma_{\lambda\lambda_2}(-P\!\!\!\!\slash+M), \NO \\
&&\sum_{\sigma_3\sigma_4}C_{\lambda\lambda_2}^{\sigma_3\sigma_4}v(p_3,\sigma_3)\overline{u}(p_4,\sigma_4)=\frac{1}{\sqrt{32}}\gamma_{\lambda\lambda_2}(P\!\!\!\!\slash+M), \NO \\
&&\sum_{\sigma_3\sigma_4}C_{\lambda\lambda_2}^{\sigma_3\sigma_4}v(p_4,\sigma_4)\overline{u}(p_3,\sigma_3)=(-1)^{\lambda+1}\frac{1}{\sqrt{32}}\gamma_{\lambda\lambda_2}(P\!\!\!\!\slash+M).
\eea

Having the above equations, the Feynman amplitude for the decay of compact tetraquarks can be formally written as
\bea
&&\langle X|T_C(P,J,J_z,+)\rangle=(\frac{2}{M})^\frac{3}{2}\Psi(\bm{0})\langle X|(cc)_{\lambda=0}(\frac{P}{2})(\bar{c}\bar{c})_{\lambda=0}(\frac{P}{2})\rangle^+, \NO \\
&&\langle X|T_C(P,J,J_z,-)\rangle=(\frac{2}{M})^\frac{3}{2}\Psi(\bm{0})\epsilon_{JJ_z}^{\mu\nu}\langle X|(cc)_{\lambda=1}(\frac{P}{2})(\bar{c}\bar{c})_{\lambda=1}(\frac{P}{2})\rangle^-_{\mu\nu}.
\eea

\subsection{Processes for inclusive hadronic decay of fully-charm tetraquarks}

The inclusive decay width of tetraquarks into light hadrons can be accessed by evaluating the following partonic processes,
\bea
&&T\rightarrow gg, \label{eqn:g} \\
&&T\rightarrow q\bar{q}, \label{eqn:q}
\eea
where $g$ and $q$ denote a gluon and a light quark, respectively,
and $q$ runs over the three light flavours, namely $u$, $d$, and $s$.
For the charmed decay, only one partonic process needs to be considered, i.e.
\bea
T\rightarrow c\bar{c}. \label{eqn:c}
\eea
Note that for spin-0 tetraquarks, process~(\ref{eqn:q}) is forbidden.

\begin{figure}
\begin{center}
\subfloat[]{\includegraphics[width=3.0cm]{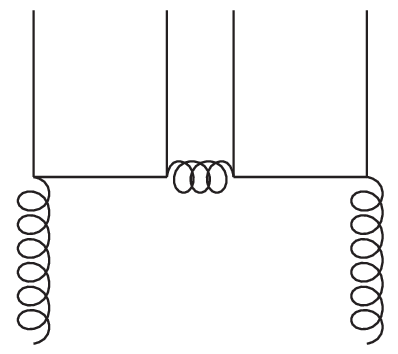} \label{diag-a}}~~~~~~
\subfloat[]{\includegraphics[width=3.0cm]{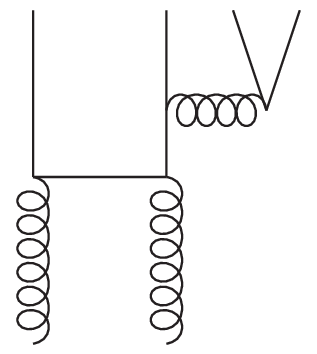} \label{diag-b}}~~~~~~
\subfloat[]{\includegraphics[width=3.0cm]{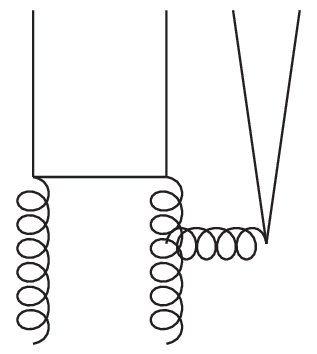} \label{diag-c}} \\
\subfloat[]{\includegraphics[width=3.0cm]{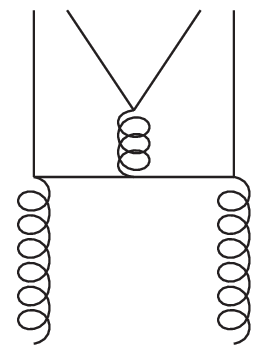} \label{diag-d}}~~~~~~
\subfloat[]{\includegraphics[width=3.0cm]{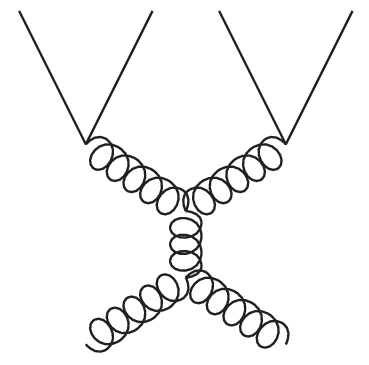} \label{diag-e}}~~~~~~
\subfloat[]{\includegraphics[width=3.0cm]{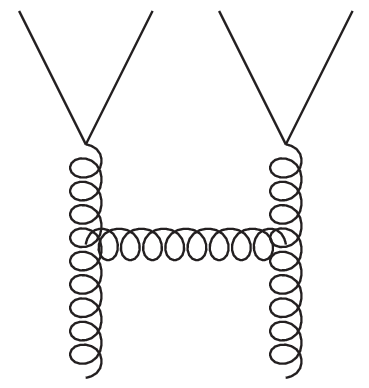} \label{diag-f}} \\
\subfloat[]{\includegraphics[width=3.0cm]{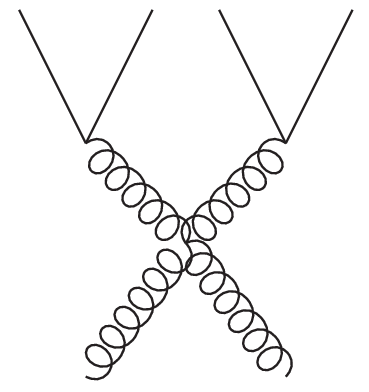} \label{diag-g}}~~~~~~
\subfloat[]{\includegraphics[width=3.0cm]{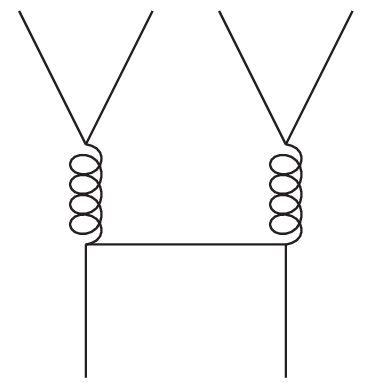} \label{diag-h}}~~~~~~
\subfloat[]{\includegraphics[width=3.0cm]{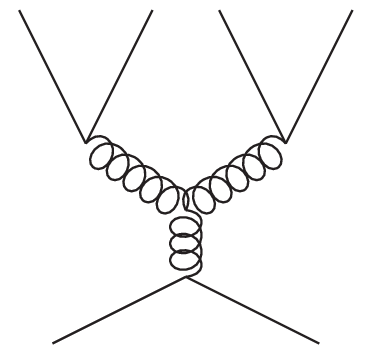} \label{diag-i}} \\
\end{center}
\caption{Topologies of the Feynman diagrams.}
\label{fig:diag}
\end{figure}

The topologies of the Feynman diagrams for process~(\ref{eqn:g}) are shown in Figures~\ref{diag-a}-\ref{diag-g}.
and Figure~\ref{diag-h} and \ref{diag-i} illstrute those for either process~(\ref{eqn:q}) or process~(\ref{eqn:c}).
The diagrams for process~(\ref{eqn:q}) can be obtained running over all possible configurations generated by the following procedure,
designating a Fermion line in Figures~\ref{diag-h} and \ref{diag-i} as a light quark line connecting the final states,
and the other two Fermion lines as charm quark lines connecting the four initial charm quarks (antiquarks) forming the tetraquark.
For process~(\ref{eqn:c}), all the Fermion lines in Figures~\ref{diag-h} and \ref{diag-i} are charm quark lines.
The diagrams can be obtained by designating a pair of charm-anticharm as final states.

Now the decay width of tetraquarks can be obtained.
The decay width of a molecule-like tetraquark to light hadrons can be written as
\bea
&&\Gamma(T_M(J)\rightarrow\mathrm{L.H.})=\Gamma(T_M(J)\rightarrow gg)+\Gamma(T_M(J)\rightarrow q\bar{q}) \NO \\
&&~~~~=\frac{1}{2M}\frac{1}{8\pi}\frac{1}{2J+1}\frac{1}{M}\left|\psi_M(\bm{0})\right|^2\sum_{J_z}\epsilon_{JJ_z}^{\mu\nu}(\epsilon_{JJ_z}^{\mu'\nu'})^* \NO \\
&&~~~~\times\left(\langle gg|J/\psi J/\psi\rangle_{\mu\nu}\langle J\psi J/\psi|gg\rangle_{\mu'\nu'}
+\sum_q\langle q\bar{q}|J/\psi J/\psi\rangle_{\mu\nu}\langle J\psi J/\psi|q\bar{q}\rangle_{\mu'\nu'}\right),
\eea
where the factor $1/(8\pi)$ is the phase space.
Similarly, its charmed decay width reads
\bea
&&\Gamma(T_M(J)\rightarrow\mathrm{C.H.})=\Gamma(T_M(J)\rightarrow c\bar{c}) \NO \\
&&=\frac{1}{2M}\frac{\sqrt{3}}{16\pi}\frac{1}{2J+1}\frac{1}{M}\left|\psi_M(\bm{0})\right|^2\sum_{J_z}\epsilon_{JJ_z}^{\mu\nu}(\epsilon_{JJ_z}^{\mu'\nu'})^*
\langle c\bar{c}|J/\psi J/\psi\rangle_{\mu\nu}\langle J\psi J/\psi|c\bar{c}\rangle_{\mu'\nu'}.
\eea
The decay widths of compact tetraquarks are presented as follows,
\bea
&&\Gamma(T_C^+(0)\rightarrow\mathrm{L.H.})=\Gamma(T_C^+(0)\rightarrow gg) \NO \\
&&~~~~~~~~=\frac{1}{2M}\frac{1}{8\pi}(\frac{2}{M})^3\left|\Psi_C^+(\bm{0})\right|^2
\langle gg|(cc)_{\lambda=0}(\bar{c}\bar{c})_{\lambda=0}\rangle^+\langle(cc)_{\lambda=0}(\bar{c}\bar{c})_{\lambda=0}|gg\rangle^+, \NO \\
&&\Gamma(T_C^+(0)\rightarrow\mathrm{C.H.})=\Gamma(T_C^+(0)\rightarrow c\bar{c}) \NO \\
&&~~~~~~~~=\frac{1}{2M}\frac{\sqrt{3}}{16\pi}(\frac{2}{M})^3\left|\Psi_C^+(\bm{0})\right|^2
\langle c\bar{c}|(cc)_{\lambda=0}(\bar{c}\bar{c})_{\lambda=0}\rangle^+\langle(cc)_{\lambda=0}(\bar{c}\bar{c})_{\lambda=0}|c\bar{c}\rangle^+, \NO \\
&&\Gamma(T_C^-(J)\rightarrow\mathrm{L.H.})=\Gamma(T_C^-(J)\rightarrow gg)+\Gamma(T_C^-(J)\rightarrow q\bar{q}) \NO \\
&&~~~~~~~~=\frac{1}{2M}\frac{1}{8\pi}\frac{1}{2J+1}(\frac{2}{M})^3\left|\Psi_C^-(\bm{0})\right|^2\sum_{J_z}\epsilon_{JJ_z}^{\mu\nu}(\epsilon_{JJ_z}^{\mu'\nu'})^* \NO \\
&&~~~~~~~~\times\Big(\langle gg|(cc)_{\lambda=1}(\bar{c}\bar{c})_{\lambda=1}\rangle^+_{\mu\nu}\langle(cc)_{\lambda=1}(\bar{c}\bar{c})_{\lambda=1}|gg\rangle^+_{\mu'\nu'} \NO \\
&&~~~~~~~~+\sum_q\langle q\bar{q}|(cc)_{\lambda=1}(\bar{c}\bar{c})_{\lambda=1}\rangle^+_{\mu\nu}\langle(cc)_{\lambda=1}(\bar{c}\bar{c})_{\lambda=1}|q\bar{q}\rangle^+_{\mu'\nu'}\Big), \NO \\
&&\Gamma(T_C^-(J)\rightarrow\mathrm{C.H.})=\Gamma(T_C^-(J)\rightarrow c\bar{c}) \NO \\
&&~~~~~~~~=\frac{1}{2M}\frac{\sqrt{3}}{16\pi}\frac{1}{2J+1}(\frac{2}{M})^3\left|\Psi_C^-(\bm{0})\right|^2\sum_{J_z}\epsilon_{JJ_z}^{\mu\nu}(\epsilon_{JJ_z}^{\mu'\nu'})^* \NO \\
&&~~~~~~~~\times\langle c\bar{c}|(cc)_{\lambda=1}(\bar{c}\bar{c})_{\lambda=1}\rangle^+_{\mu\nu}\langle(cc)_{\lambda=1}(\bar{c}\bar{c})_{\lambda=1}|c\bar{c}\rangle^+_{\mu'\nu'},
\eea
where the superscripts, $+$ and $-$, are to indicate the associated colour functions.
Here we do not distinguish wave functions for different total angular momentum in the spirit of the heavy quark spin symmetry.

\section{Phenomenological Results\label{sec:result}}

The analytical results for the decay widths are presented below.
\bea
&&\Gamma(T_M(0)\rightarrow gg)=\frac{32,768\pi^3\alpha_s^4}{27M^8}\left|\psi_M(\bm{0})\right|^2\left|\psi_{J/\psi}(\bm{0})\right|^4, \NO \\
&&\Gamma(T_M(0)\rightarrow c\bar{c})=\frac{11,075,584\sqrt{3}\pi^3\alpha_s^4}{243M^8}\left|\psi_M(\bm{0})\right|^2\left|\psi_{J/\psi}(\bm{0})\right|^4, \NO \\
&&\Gamma(T_M(2)\rightarrow gg)=\frac{25,690,112\pi^3\alpha_s^4}{405M^8}\left|\psi_M(\bm{0})\right|^2\left|\psi_{J/\psi}(\bm{0})\right|^4, \NO \\
&&\Gamma(T_M(2)\rightarrow q\bar{q})=\frac{8,388,608\pi^3\alpha_s^4}{135M^8}n_l\left|\psi_M(\bm{0})\right|^2\left|\psi_{J/\psi}(\bm{0})\right|^4, \NO \\
&&\Gamma(T_M(2)\rightarrow c\bar{c})=\frac{719,323,136\sqrt{3}\pi^3\alpha_s^4}{1,215M^8}\left|\psi_M(\bm{0})\right|^2\left|\psi_{J/\psi}(\bm{0})\right|^4, \NO \\
&&\Gamma(T_C^+(0)\rightarrow gg)=\frac{991,232\pi^3\alpha_s^4}{27M^8}\left|\Psi_C^+(\bm{0})\right|^2, \NO \\
&&\Gamma(T_C^+(0)\rightarrow c\bar{c})=\frac{16,384\sqrt{3}\pi^3\alpha_s^4}{27M^8}\left|\Psi_C^+(\bm{0})\right|^2, \NO \\
&&\Gamma(T_C^-(0)\rightarrow gg)=\frac{147,456\pi^3\alpha_s^4}{M^8}\left|\Psi_C^-(\bm{0})\right|^2, \NO \\
&&\Gamma(T_C^-(0)\rightarrow c\bar{c})=\frac{3,964,928\sqrt{3}\pi^3\alpha_s^4}{81M^8}\left|\Psi_C^-(\bm{0})\right|^2, \NO \\
&&\Gamma(T_C^-(2)\rightarrow gg)=\frac{12,845,056\pi^3\alpha_s^4}{135M^8}\left|\Psi_C^-(\bm{0})\right|^2, \NO \\
&&\Gamma(T_C^-(2)\rightarrow q\bar{q})=\frac{4,194,304\pi^3\alpha_s^4}{45M^8}n_l\left|\Psi_C^-(\bm{0})\right|^2, \NO \\
&&\Gamma(T_C^-(2)\rightarrow c\bar{c})=\frac{359,661,568\sqrt{3}\pi^3\alpha_s^4}{405M^8}\left|\Psi_C^-(\bm{0})\right|^2,
\eea
where $n_l$ is the number of light flavours.
For fully-charm tetraquark decay, $n_l=3$,
while for fully-bottom tetraquark decay, $n_l=4$.
Note that our results for $T_C\rightarrow\mathrm{L.H.}$ agree with those in Reference~\cite{Sang:2023ncm}.

Although there have been numerous works solving the tetraquark systems,
their results on the values of the wave functions at the origin diverge,
which is due to lack of data.
To this end, we are only concerned with the normalized short-distance coefficients ($\hat{\Gamma}$) in this paper.

Before we present further results, it is important to note that $T_C^+(0)$ and $T_C^-(0)$ have completely identical quantum numbers,
thus, they are physically indistinguishable.
The physical states should be two orthogonal mixtures of them.
Assuming the mixing angle for one of the physical states is $\vartheta$,
the Feynman amplitude for its decay is
\bea
\langle X|T_C(0)\rangle=\langle X|T_C^+(0)\rangle\mathrm{cos}\vartheta+\langle X|T_C^-(0)\rangle\mathrm{sin}\vartheta.
\eea
Up to QCD leading order, the short-distance part for either of the amplitudes on the right-hand side of the above equation is real,
therefore the interference contribution to the decay width is twice of the geometric mean of the decay widths for colour symmetric and asymmetric states, namely
\bea
\Gamma(T_C^{inter}(0)\rightarrow X)=2\sqrt{\Gamma(T_C^+(0)\rightarrow X)\Gamma(T_C^-(0)\rightarrow X)}, \label{eqn:inter}
\eea
and the decay widths of the spin-0 state with mixing angle $\vartheta$ is
\bea
&&\Gamma(T_C(0)\rightarrow X) \NO \\
&&~~=\Gamma(T_C^+(0)\rightarrow X)\mathrm{cos}^2\vartheta+\Gamma(T_C^{inter}(0)\rightarrow X)\mathrm{cos}\vartheta\mathrm{sin}\vartheta
+\Gamma(T_C^-(0)\rightarrow X)\mathrm{sin}^2\vartheta.\label{eqn:mix}
\eea

However, the two decay widths on the right-hand side of Equation~(\ref{eqn:inter}) involve different wave functions, of which the values are still unknown,
we do not concern ourselves with the interference contribution and present results only for the unphysical states with specific colour configurations.
Once the wave functions are solved, one can employ Equation~(\ref{eqn:inter}) and Equation~(\ref{eqn:mix}) to obtain the interference contributions.

With the results given above, it is easy to obtain the following analytical results for the decay widths,
which is presented in the form of normalized short-distance coefficient.
One can obtain the decay width by multiplying the corresponding squared wave function at the origin.
\bea
&&\hat{\Gamma}(T_M(0)\rightarrow\mathrm{L.H.})=\frac{32,768\pi^3\alpha_s^4}{27M^8}, \NO \\
&&\hat{\Gamma}(T_M(0)\rightarrow\mathrm{C.H.})=\frac{11,075,584\sqrt{3}\pi^3\alpha_s^4}{243M^8}, \NO \\
&&\hat{\Gamma}(T_M(2)\rightarrow\mathrm{L.H.})=\frac{\pi^3\alpha_s^4}{M^8}(\frac{25,690,112}{405}+\frac{8,388,608}{135}n_l), \NO \\
&&\hat{\Gamma}(T_M(2)\rightarrow\mathrm{C.H.})=\frac{719,323,136\sqrt{3}\pi^3\alpha_s^4}{1,215M^8}, \NO \\
&&\hat{\Gamma}(T_C^+(0)\rightarrow\mathrm{L.H.})=\frac{\pi^3\alpha_s^4}{M^8}\frac{991,232}{27}, \NO \\
&&\hat{\Gamma}(T_C^+(0)\rightarrow\mathrm{C.H.})=\frac{\pi^3\alpha_s^4}{M^8}\frac{16,384\sqrt{3}}{27}, \NO \\
&&\hat{\Gamma}(T_C^-(0)\rightarrow\mathrm{L.H.})=\frac{\pi^3\alpha_s^4}{M^8}147,456, \NO \\
&&\hat{\Gamma}(T_C^-(0)\rightarrow\mathrm{C.H.})=\frac{\pi^3\alpha_s^4}{M^8}\frac{3,964,928\sqrt{3}}{81}, \NO \\
&&\hat{\Gamma}(T_C^-(2)\rightarrow\mathrm{L.H.})=\frac{\pi^3\alpha_s^4}{M^8}(\frac{12,845,056}{135}+\frac{4,194,304}{45}n_l), \NO \\
&&\hat{\Gamma}(T_C^-(2)\rightarrow\mathrm{C.H.})=\frac{359,661,568\sqrt{3}\pi^3\alpha_s^4}{405M^8}.
\eea

For each tetraquark state, it is interesting to investigate the ratio of the decay width for the light-hadron final states to that for the charmed final states, namely
\bea
\mathcal{R}(T)\equiv\frac{\Gamma(T\rightarrow\mathrm{L.H.})}{\Gamma(T\rightarrow\mathrm{C.H.})}.
\eea
On the theory side, this ratio is a number independent of physical quantities such as hadron mass and coupling constants,
and is thus free of uncertainties coming from parameter choices.
On the experiment side, compared to the measurement of the decay width,
it is much easier to measure the ratio thanks to the cancellation of the systematic uncertainties.

The values of $\mathcal{R}$ are obtained as
\bea
&&\mathcal{R}(T_M(0))=\frac{3\sqrt{3}}{338}\approx0.0154, \NO \\
&&\mathcal{R}(T_M(2))=\frac{\sqrt{3}}{28}+\frac{12\sqrt{3}}{343}n_l\approx0.0619+0.0606n_l, \NO \\
&&\mathcal{R}(T_C^+(0))=\frac{121\sqrt{3}}{6}\approx34.9, \NO \\
&&\mathcal{R}(T_C^-(0))=\frac{243\sqrt{3}}{242}\approx1.74, \NO \\
&&\mathcal{R}(T_C^-(2))=\frac{\sqrt{3}}{28}+\frac{12\sqrt{3}}{343}n_l\approx0.0619+0.0606n_l.
\eea
For fully-charm tetraquarks,
\bea
\mathcal{R}(T_M(2))=\mathcal{R}(T_C^-(2))=\frac{193\sqrt{3}}{1372}\approx0.244,
\eea
while for fully-bottom tetraquarks,
\bea
\mathcal{R}(T_M(2))=\mathcal{R}(T_C^-(2))=\frac{241\sqrt{3}}{1372}\approx0.304.
\eea

For both of the cases, the ratios $\mathcal{R}$ for the spin-2 tetraquarks are much different from those for the spin-0 ones,
which can serve as an alternative way of determining the spin of a fully-heavy tetraquark candidate,
in addition to measuring the angular distribution of the final-state lepton pairs in the decay process
$T\rightarrow J/\psi J/\psi\rightarrow l^+l^-l^+l^-$.
For spin-0 tetraquarks, $\mathcal{R}$ is different for different colour configurations or binding mechanisms.
Note that such information is crucial for understanding the binding mechanism of quarks,
however is not accessible by most of the observations.

\section{Summary\label{sec:summary}}

In this paper, we presented the theoretical framework of calculating the hadronic decay width of fully-heavy tetraquarks.
Both molecule-like states and genuine tetraquarks were taken into consideration.
The phenomenological results at QCD leading order were presented,
and a crucial physical quantities, say the ratio $\mathcal{R}$ of the decay width for the light-hadron final states to that for the charmed final states were also evaluated.
This value is a number independent of all the physical parameters,
and exhibits good properties for exploring the nature of the observed double charmonia resonances:
it can not only help to determine the spin of the resonances, but also distinguish different colour configurations and binding mechanisms of the spin-0 tetraquarks.
It is expected that our results may help the experiment to acquire useful information that is not accessible by any other tools.

\begin{acknowledgments}
This work is supported by the National Natural Science Foundation of China (Grant No. 11965006),
and the Guizhou Provincial Science and Technology Project–9–under the Grant No. QKH-Basic-ZK[2021]YB319.
Y.-P. Y. acknowledges support from SUT and the Office of the Higher Education Commission under the NRU project of Thailand.
\end{acknowledgments}

\providecommand{\href}[2]{#2}\begingroup\raggedright\endgroup

\end{document}